\documentclass[prc,twocolumn,superscriptaddress,showpacs,twoside,floatfix]
{revtex4-1}

\usepackage{amssymb,epsfig}

\begin{document}

\title{Decay mechanism and lifetime of $^{67}$Kr}

\author{L.V.~Grigorenko}
\affiliation{Flerov Laboratory of Nuclear Reactions, JINR,  RU-141980 Dubna,
Russia}
\affiliation{National Research Nuclear University ``MEPhI'', Kashirskoye shosse
31, RU-115409 Moscow, Russia}
\affiliation{National Research Centre ``Kurchatov Institute'', Kurchatov sq.\ 1,
RU-123182 Moscow, Russia}
\author{T.A.~Golubkova}
\affiliation{Advanced Educational and Scientific Center, Moscow State
University, Kremenchugskaya 11, RU-121357 Moscow, Russia}
\author{J.S.~Vaagen}
\affiliation{Institute of Physics and Technology, University of Bergen, N-5007 Bergen, Norway}
\author{M.V.~Zhukov}
\affiliation{Department of Physics, Chalmers University of Technology, S-41296
G\"{o}teborg, Sweden}


\begin{abstract}
The lifetime of the recently discovered $2p$ emitter $^{67}$Kr was recently found considerably below the lower limit predicted theoretically. This communication addresses this issue. Different separation energy systematics are analyzed and different mechanisms for $2p$ emission are evaluated. It is found that the most plausible reason for this disagreement is a decay mechanism of $^{67}$Kr, which is not ``true $2p$'' emission, but ``transition dynamics'' on the borderline between true $2p$ and sequential $2p$ decay mechanisms. If this is true, this imposes stringent limits $E_r=1.35-1.42$ MeV on the ground state energy of $^{66}$Br relative to the $^{65}$Se-$p$ threshold.
\end{abstract}


\maketitle


\textit{Introduction.}
%
%
---  Discovery of a new case of two-proton ($2p$) radioactive decay has been reported recently, in $^{67}$Kr \cite{Goigoux:2016}. This is an important advance in the field as $^{67}$Kr is the heaviest $2p$ emitter observed so far providing new opportunities for refining our understanding of the $2p$ radioactivity phenomenon. For relatively small $2p$ decay energy (found to be $E_T=1690 \pm 17$ keV) the $2p$ decay branch is in a tough competition with weak transitions, providing only $37(14)\%$ of the decay probability. The measured total and partial $2p$ lifetimes are $7.4(30)$ ms and $20(11)$ ms respectively. The previous (2003) theoretical predictions within the three-body cluster model from Ref.\ \cite{Grigorenko:2003b} provided a \emph{lower} lifetime limit of $240^{+100}_{-70}$ ms (within experimental decay energy uncertainty). Here, we are going to make refined theoretical calculations and discuss possible origins of the observed discrepancy.

Two-proton radioactivity predicted in 1960 \cite{Goldansky:1960} was experimentally discovered in $^{45}$Fe in 2002 \cite{Pfutzner:2002,Giovinazzo:2002} after four decades of dedicated search. Since that time our knowledge in this field has expanded tremendously. Radioactive $2p$ decay has been found in $^{19}$Mg, $^{48}$Ni, and $^{54}$Zn and later first information about two-proton correlations was obtained for these isotopes \cite{Pfutzner:2012}. Highly precise information about three-body correlations was obtained for $^{6}$Be \cite{Grigorenko:2009c,Egorova:2012,Fomichev:2012} and $^{16}$Ne \cite{Brown:2014,Brown:2015a}. Evidence has been obtained that $2p$ decay of $^{30}$Ar is of the ``transition type'' with decay mechanism on the borderline between true $2p$  decay and sequential $2p$ decay \cite{Mukha:2015,Golubkova:2016}.

The theoretical description of the $2p$ emission based on the three-cluster core+$p$+$p$ model \cite{Grigorenko:2000b} has achieved a high level of sophistication \cite{Pfutzner:2012}. The lifetimes for the number of $2p$ emitters were successfully predicted \cite{Grigorenko:2003b,Pfutzner:2012}. The predicted connections between configuration mixing in the wave function (WF) structure and correlations in $2p$ decay were experimentally confirmed \cite{Miernik:2007b,Grigorenko:2009}. Detailed comparison of the theoretical and observed correlations for the lightest $2p$ emitters $^{6}$Be and $^{16}$Ne demonstrated high accuracy of the approach \cite{Grigorenko:2009c,Egorova:2012,Brown:2014,Brown:2015a}. We draw attention to the studies \cite{Brown:2014}, which demonstrated the ability to describe very delicate long-range effects of the three-body Coulomb continuum problem. Thus the predictions of the three-body approach were so far reliable and therefore it is important to understand what went wrong in the $^{67}$Kr case.

There are several possible reasons why the lifetime could have been overestimated in the three-body decay calculations \cite{Grigorenko:2003b}:

\noindent (i) Slow convergence was demonstrated for three-body calculations when the two-body resonance energy $E_r$ becomes sufficiently low and enters the two-proton decay energy window from above: $E_r>0.84 E_T$, $E_r \rightarrow 0.84 E_T$. This problem was overcome in the later (2007) studies Ref.\ \cite{Grigorenko:2007,Grigorenko:2007a} focusing on $^{17}$Ne and $^{45}$Fe cases. However, the predictions for heavier systems were not revised.

\noindent (ii) The ``standard'' systematics of potential radii $a = r_0 (A_{\text{core}}+1)^{1/3}$, where $r_0 = 1.2$ fm,  was used in the pioneering studies \cite{Grigorenko:2003b}. For a nuclear system as heavy as $^{67}$Kr, lifetime sensitivity to this choice could be more important than for the lighter systems. Some additional adjustment of the single-particle orbital properties of the calculations is highly desirable.

\noindent (iii) Logical continuation of the trend of item (i) is further to decrease the $E_r$ value to the range $0.8 E_T < E_r < 0.84 E_T$. The phenomenon which can drastically increase the width and thus decrease the lifetime of a $2p$ emitter is a transition from true $2p$ to sequential $2p$ decay mechanism. Our studies have shown that this question is especially important for $^{67}$Kr as various estimate results point to an energy region for $E_r$ which does not allow to exclude such a possibility.

\begin{figure*}
\includegraphics[width=0.392\textwidth]{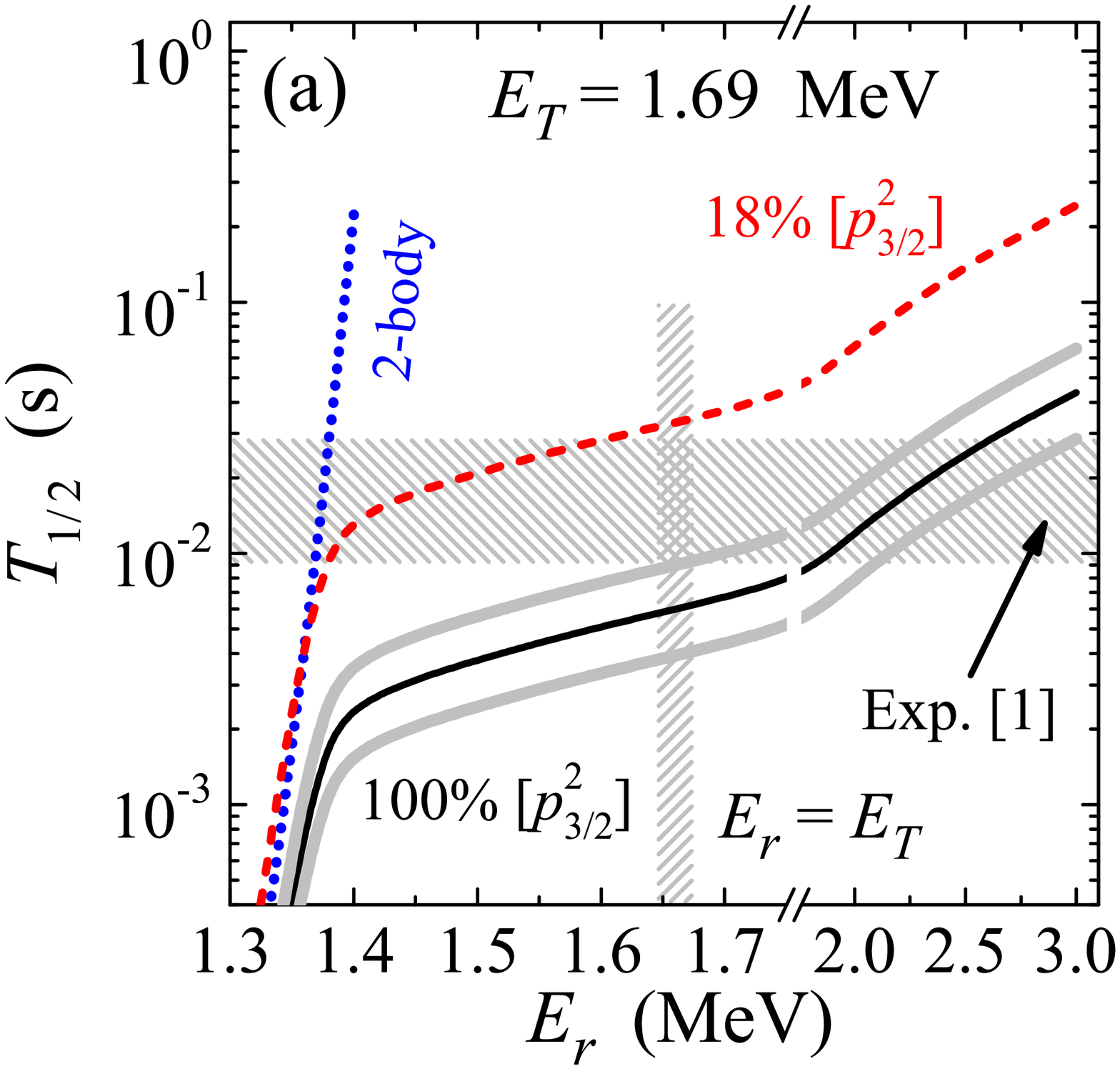}
\includegraphics[width=0.321\textwidth]{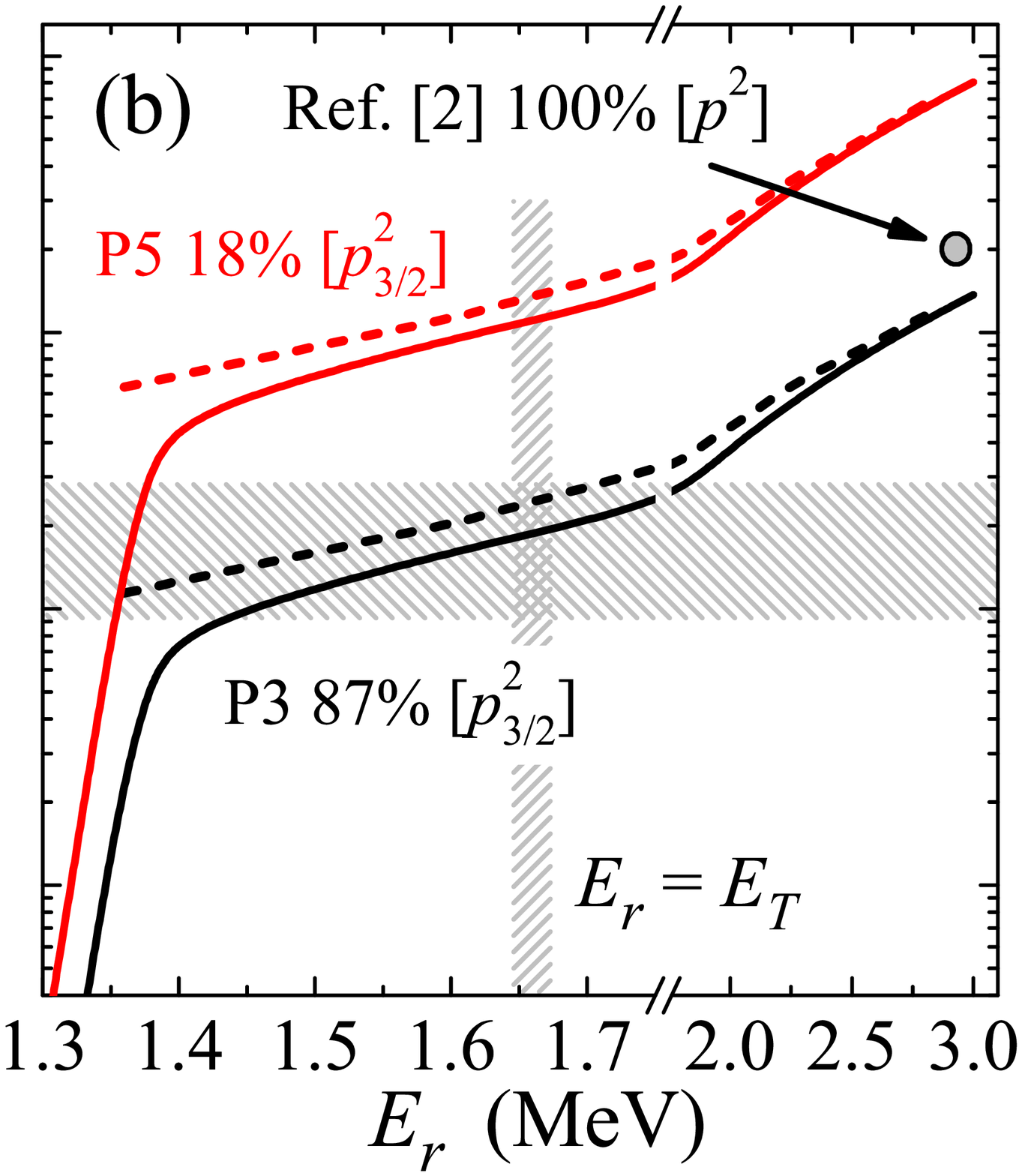}

\caption{(Color online) Lifetime of $^{67}$Kr as a function of the $p$-wave ground state resonance energy $E_r$ in $^{66}$Br for fixed energy $E_T=1.69$ MeV. (a) IDDM model. Solid black and dashed red curves correspond to different assumed weights of $[p^2]$ configuration. Solid gray curves corresponds to solid black curve with $\pm 0.2$ fm modified channel radius $r_{\text{cp}}$.  The blue dotted line shows the two-body R-matrix estimate of decay width into $^{66}$Br+$p$ channel with energy $E_T-E_r$. (b) Dashed lines show three-body model results for P3 and P5 potential producing different configuration mixing. Solid lines show extrapolation of three-body results from large $E_r$ values (where they are converged) by IDDM curves.}
\label{fig:lifetime-ot-er}
\end{figure*}

In this work we repeated calculations \cite{Grigorenko:2003b} for $^{67}$Kr with technical improvements. We performed separation energy evaluations, introduced better restricted core-$p$ interactions, and performed the $^{67}$Kr lifetime studies both in R-matrix semianalytical and full three-body decay models. We conclude that variant (iii) seems to be the most plausible explanation of the situation.


\textit{Width estimates within improved direct decay model.}
%
%
--- The semianalytical R-matrix \emph{direct decay model} of $2p$ decay is a convenient tool for simple evaluation of $2p$ lifetimes and systematic studies \cite{Grigorenko:2007,Grigorenko:2007a,Pfutzner:2012}. Thus
\begin{eqnarray}
\frac{d \Gamma_{j_1 j_2}(E_T)}{d \varepsilon} =  \frac{E_{T}\left \langle V_{3}\right \rangle
^{2}} {2\pi} f^{(J)}_{j_1 j_2} \,
 \frac{ \Gamma_{j_1}(E_1)}
{(E_{1}-E_{j_1})^{2}+\Gamma_{j_1}(E_1)^{2}/4}
\nonumber \\
 \times \, \frac{\Gamma_{j_2}(E_{2})}
{(E_{2}-E_{j_2})^{2} + \Gamma_{j_2}
(E_{2})^{2}/4}\;,\qquad
\label{eq:sequent}
\end{eqnarray}
where $E_{j_i}$ are two-body resonance energies in core-$p$ subsystem number $i=\{1,2\}$, $\varepsilon$ denotes an energy distribution parameter, $0\leq \varepsilon \leq 1$, $E_1=\varepsilon E_T$, $E_2=(1-\varepsilon) E_T$, and $\Gamma_{j_i}$ is the standard R-matrix expression for the width as a function of the energy for the involved resonances in the core+$p$ subsystems. The matrix element $\left \langle V_{3}\right \rangle$ can be well approximated
as
\[
\left \langle V_{3}\right \rangle ^{2}   =   D_3 [(E_T-E_{j_1}-E_{j_2})^2 +
(\Gamma_{j_1}(E_{j_1})+\Gamma_{j_2}(E_{j_2}))^2/4]\, ,
\]
where the dimensionless parameter $D_3 \approx 1.0-1.5$ is a constant. The structure factor $f^{(J)}_{j_1 j_2}$ describes the weight of the $[j_1 j_2]_J$ configuration in the nuclear interior WF.

The direct decay model works well when decay via a single quantum configuration is dominating with two nucleons being emitted from states with definite single-particle angular momenta $j_1$ and $j_2$ sharing the total decay energy $E_T$. The effect of the low-lying states of $^{66}$Br on the width of $^{67}$Kr can be well evaluated in such a model. We actually use the \emph{improved direct decay model}  (IDDM) of \cite{Golubkova:2016}, which utilizes a more complex semianalytical approximation and is more phenomenologically tuned to the phenomenon. The IDDM lifetimes of $^{67}$Kr are shown  in Fig.\ \ref{fig:lifetime-ot-er} (a) as a function of the $^{66}$Ga ground state (g.s.) energy. The g.s.\ of $^{66}$Ga has $J^{\pi}=0^+$. This means that it is likely to involve $p_{3/2}$ single-particle state coupled to the $^{65}$Se g.s.\ with $J^{\pi}=3/2^-$. It is also very likely that this will be the $^{66}$Br g.s.\ because of strong Thomas-Ehrman effect. So, we assume $[p^2_{3/2}]_0$ for the $^{67}$Kr g.s.\ decay and $E_1=E_2=E_r$. The width calculations for such a heavy $2p$ emitter are very sensitive to the exact properties of the Coulomb barrier. The R-matrix channel radius $r_{\text{cp}}=6.12$ fm and reduced width $\theta^2=1$ are chosen to reproduce exactly the width of the resonance obtained with potential P1 (discussed below). Small variation of the value $r_{\text{cp}}$ by $\pm 0.2$ fm, leads to more than a factor 2 variation of the lifetime, see gray curves in Fig.\ \ref{fig:lifetime-ot-er} (a). It is predicted in shell-model calculations of Ref.\ \cite{Goigoux:2016} that the weight of the $[p^2_{3/2}]$ configuration in $^{67}$Kr is only $\sim 18 \%$. This may imply a corresponding increase of the lifetime predicted in IDDM (red dashed curve).

Two conclusions can be made from these calculations. (a) The transition ``true $2p$'' $\rightarrow$ ``sequential $2p$'' is taking place in $^{67}$Kr in the $E_r=1.35-1.42$ MeV range ($S_p$ from $-340$ to $-270$ keV). (b) For pure $[p^2_{3/2}]$ structure of $^{67}$Kr the calculated lifetime is consistent with experiment for quite broad range of $E_r \sim 1.7-2.7$ MeV. In contrast, for realistic shell model structure predicted in \cite{Goigoux:2016} ($\sim 18 \%$ of $[p^2_{3/2}]$), the calculated lifetimes agree with experiment only in much narrower range $E_r \sim 1.38-1.58$ MeV strongly overlaping with ``transition dynamics'' range. Below we try to understand how realistic the latter possibility is from other points of view as well.

\begin{table}[b]
\caption{The first column shows the $S_p$ range for transition ``true $2p$'' $\rightarrow$ ``sequential $2p$'' decay mechanisms in $^{67}$Kr. The other columns show results of different systematic evaluations of the upper and lower boundaries for $S_p$ value in $^{67}$Kr.}
\begin{ruledtabular}
\begin{tabular}[c]{cccccc}
  & transition & Ref.\ \cite{Ormand:1997} & $S_p/S_n$  & OES & TES \\
\hline
upper  & $-290$ & 130  & 180  & 200  & $-80$   \\
lower  & $-340$ & $-440$ & $-120$ & $-340$ & $-320$  \\
\end{tabular}
\end{ruledtabular}
\label{tab:ranges}
\end{table}

\begin{table}[b]
\caption{Parameters of Woods-Saxon potentials with surface (ls) interaction in the $^{65}$Se-$p$ channel. Energies are in MeV, distances in fm.}
\begin{ruledtabular}
\begin{tabular}[c]{cccccccc}
Pot. & l & $V_c$ & $a$ & $d$ & $V_{ls}$ & $r_{\text{sph}}$ & $r_{\text{ch}}$  \\
\hline
P1  & 1  & -50.485 & 4.85   & 0.65  & 1.5  & 5.5   & 4.21 \\
P2  & 1  & -20.89  & 4.825  & 0.65  & 0.5  & 6.248 & 4.82 \\
\end{tabular}
\end{ruledtabular}
\label{tab:pot12}
\end{table}

\begin{table}[b]
\caption{Woods-Saxon potentials with volume $(ls)$ interaction and repulsive core in the $^{65}$Se-$p$ channel. Potential P5 has the same $l=1$ and Coulomb components as P3, so we give only $l=3$ part of this potential.}
\begin{ruledtabular}
\begin{tabular}[c]{ccccccccc}
\multicolumn{2}{c}{Pot.} & l & $V_1$ & $a_1$ & $d_1$  & $V_2$ & $a_2$ & $d_2$ \\
\hline
P3 & c  & 1  & -26.389 & 5.0  & 0.65  & 75  & 1.5  & 0.53   \\
   & ls & 1  & -1.0 &  5.0 & 0.65 & \multicolumn{3}{c}{$r_{\text{sph}}=5.5$ fm} \\
P4 & c  & 1  & -57.612 & 4.55  & 0.65  & 150  & 2.7  & 0.53   \\
   & ls & 1  & -0.2 &  4.55   &  0.65 & \multicolumn{3}{c}{$r_{\text{sph}}=4.252$ fm} \\
P5 & c  & 3  & -45.8  & 5.0   & 0.65  &     &     &      \\
   & ls & 3  & 0.2    & 4.55  &  0.65 &     &     &
\end{tabular}
\end{ruledtabular}
\label{tab:pot34}
\end{table}

\begin{table}[b]
\caption{The widths $\Gamma_r$ (in eV) of the $p_{3/2}$ g.s.\ resonance of $^{66}$Br at $E_r=1.36$ MeV obtained with different $^{65}$Se-$p$ potentials. The depth parameter for P2 potential was modified to give the same resonance energy.}
\begin{ruledtabular}
\begin{tabular}[c]{cccccc}
Pot. & P1 & P2 & P3 & P4 & P5   \\
\hline
$\Gamma_r$   & 1.18  & 0.44 & 0.74   & 1.13  & 0.74 \\
\end{tabular}
\end{ruledtabular}
\label{tab:widths}
\end{table}

\begin{figure}
\includegraphics[width=0.43\textwidth]{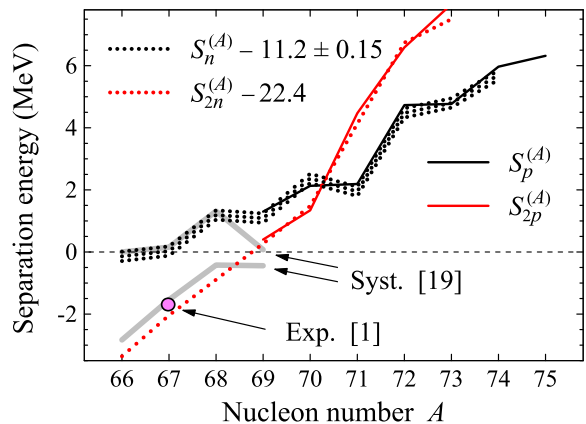}
\caption{(Color online) Systematics of $N$ and $2N$ separation energies ($S^{(A)}_N$ and $S^{(A)}_{2N}$, respectively) for Krypton isobar and mirror isotone members with mass number $A$. The isotone values are provided with constant offsets (in MeV) for visual comparison. Thick gray lines show systematics predictions from Ref.\ \cite{Ormand:1997}.}
\label{fig:en-syst-sep}
\end{figure}


\textit{Systematic consideration of separation energies.}
%
%
--- We have pointed above that it is necessary to clarify the decay mechanism of $^{67}$Kr in order to clarify the problem of the $^{67}$Kr lifetime. For the decay mechanism the question about the relation of $p$ and $2p$ separation energies $S_p$ and $S_{2p}$ is decisive \cite{Golubkova:2016}. Here we try three different types of estimates for these values.

Fig.\ \ref{fig:en-syst-sep} shows the systematics of $N$ and $2N$ separation energies for Krypton isobar and mirror isotone. The energy trends of the isotone and isobar nicely overlap in the mass range, where both of them are experimentally known. Extrapolation provided for $S_{2p}$ agrees very well with data \cite{Goigoux:2016} and systematics studies \cite{Ormand:1997}. The extrapolation for $S_p$ seems to be more uncertain and points to $S_p=30\pm 150$ keV (also in a good agreement with \cite{Ormand:1997}). Here we should keep in mind that Thomas-Ehrman shift (TES) can easily modify this value to more negative values (see discussion in \cite{Grigorenko:2015,Mukha:2015}): $S_p= -290$ keV which is just $200-300$ keV of extra binding is needed to get into the transition regime.

\begin{figure}
\includegraphics[width=0.43\textwidth]{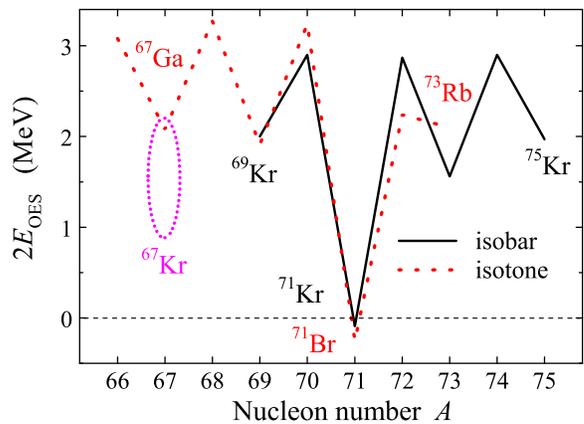}
\caption{(Color online) Systematics of odd-even staggering energies $E_{\text{OES}}$ for Krypton isobar and mirror isotone. Pink dotted ellips shows the expected uncertainty for $^{67}$Kr due to the Thomas-Ehrman effect.}
\label{fig:en-syst-oes}
\end{figure}

Fig.\ \ref{fig:en-syst-oes} shows the systematics of experimental odd-even staggering energies $2E_{\text{OES}}=S^{(A)}_{2N}-2S^{(A-1)}_{N}$ for Krypton isobar and mirror isotone. If we use the extrapolated value  $2E_{\text{OES}}=2$ MeV then $S_p=200$ keV is obtained. However, it is known that the TES effect strongly changes this systematics (again, see \cite{Grigorenko:2015,Mukha:2015}). If we take $2E_{\text{OES}}=1$ MeV then $S_p=-340$ keV is obtained, which corresponds to decay dynamics deeply in the transition region.

\begin{figure}
\includegraphics[width=0.43\textwidth]{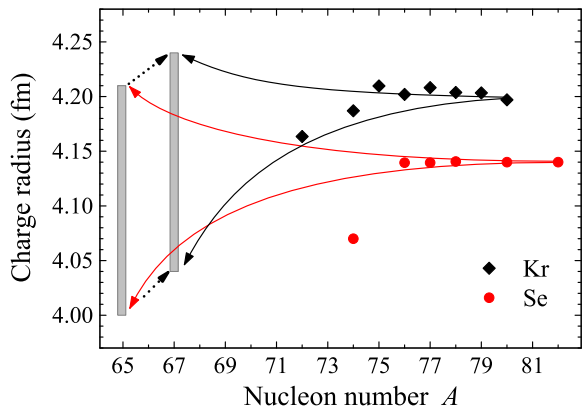}
\caption{(Color online) Systematics of charge radii for Krypton and Selenium isobars. Solid arrows extrapolate rising and falling trends along the isotopic chains which could be expected basing on data on other nuclei. Gray rectangle shows expected uncertainty for $^{65}$Se. The dashed arrows translate it into $^{67}$Kr charge radius uncertainty by using an independent particle model, which seems consistent with analogous extrapolation for the Krypton isobaric chain.}
\label{fig:en-syst-rad}
\end{figure}

The third estimate is a direct $S_p$ evaluation in a potential model which contains the Thomas-Ehrman effect. For this estimate we need two main ingredients: potential radius and charge radius of the core. We take a Woods-Saxon potential with ``standard'' systematic parameters (P1, see Table \ref{tab:pot12}) $a=r_0(A_{\text{core}}+1)^{1/3}=4.85$ fm and diffuseness $d=0.65$ fm. The experimental systematics of charge radii \cite{Angeli:2013} for Krypton and Selenium isotopes is shown in Fig.\ \ref{fig:en-syst-rad}. We can expect both falling and rising trends when approaching the proton dripline. Thus we take a relatively broad range $r_{\text{ch}}=4.0-4.21$ fm for $^{65}$Se. The corresponding Coulomb potential of a homogeneously charged sphere is used with radius $r_{\text{sph}}$ defined as
\[
r^2_{\text{sph}}=5r^2_{\text{ch}}/3+0.8^2 \,.
\]
The $S_p$  calculated in a single-particle potential model based on this uncertainty of the charge radius, ranges from $S_p=-320$ keV to $S_p=-80$ keV.

The results of all estimates are summarized in Table \ref{tab:ranges}. We see here that various approaches provide $S_p$ values with uncertainties which does not exclude possibility of the ``transition type'' decay mechanism.


\textit{Three-body decay calculations.}
%
%
--- The cluster core+$p$+$p$ model of $2p$ radioactivity is based on solution of a three-body Schr\"odinger equation with complex energy and pure outgoing wave boundary conditions, within the hyperspherical harmonics method:
\begin{eqnarray}
(\hat{H}_3-E_T + i \Gamma/2 ) \Psi^{(+)}_{E_T} = 0 \, ,  \nonumber \\
\hat{H}_3 = \hat{T}_3 + V_{p_1\text{-}p_2} + V_{\text{core-}p_1} + V_{\text{core-}p_2} + V_3(\rho) \, ,
\label{eq:sch}
\end{eqnarray}
where $\hat{T}_3$ is the three-body kinetic energy, $V_{ij}$ are  pairwise interactions between clusters, chosen based on available experimental information, and $V_3(\rho)$ is a phenomenological short-range potential depending only on a collective variable (hyperradius $\rho$) which is used to tune the total decay energy in the systematic lifetime calculations. The three-body decay width $\Gamma$ is defined using a so-called ``natural'' definition of the width $\Gamma = j/N$. Here $j$ is outgoing flux via the hypersphere of some large radius associated with WF $\Psi^{(+)}_{E_T}$, while $N$ is normalization of the WF $\Psi^{(+)}_{E_T}$ inside this sphere.

\begin{figure}
\includegraphics[width=0.43\textwidth]{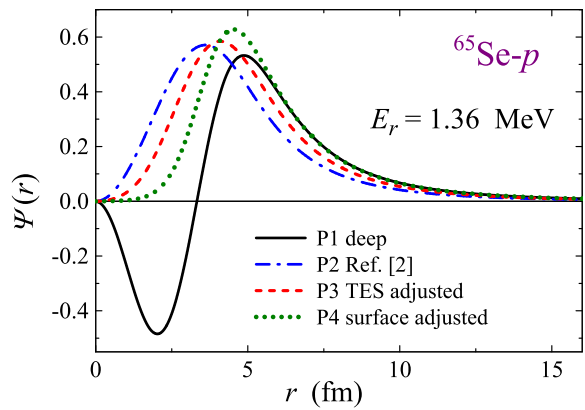}
\caption{(Color online) Resonance WFs in the $^{65}$Se-$p$ channel for different considered potentials. The potentials for all calculations were adjusted to give a resonance energy in $^{66}$Br $E_r=1.36$ MeV.}
\label{fig:res-wfs}
\end{figure}

In the $p$-$p$ channel we use semirealistic nucleon-nucleon potential Ref.\ \cite{Gogny:1970}. The employed version of the HH method works with potentials without forbidden states. For that reason, when we turn to core-$p$ potentials, we need some substitute for potential P1 (Table \ref{tab:pot12}) used in the TES estimates above. In paper \cite{Grigorenko:2003b} the potential P2 was used. It was utilized together with Coulomb potential of charged sphere with radius $r_{\text{sph}}=6.25$ fm, which means an unrealistic charge radius and wrong TES systematics. For this work we produced potential sets P3--P5, see Table \ref{tab:pot34} and Fig.\ \ref{fig:res-wfs}. Potential P3 was constructed to reproduce the systematics of the TES for potential P1 which means that this WF has analogous average orbital size. Potential P4 has the same behavior as P1 in the surface region and therefore practically the same resonance decay width, see  Fig.\ \ref{fig:res-wfs} and Table \ref{tab:widths}.

\begin{figure}
\includegraphics[width=0.45\textwidth]{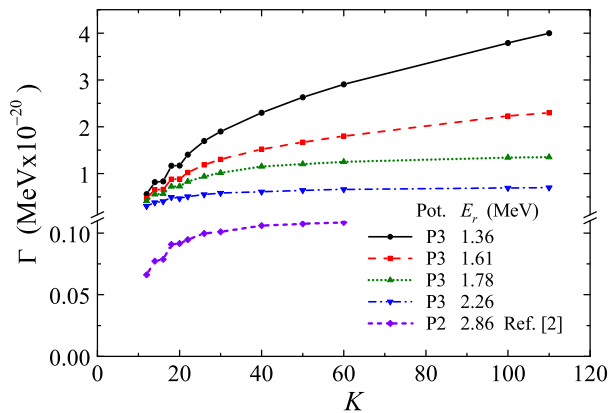}
\caption{(Color online) Convergence of the three-body lifetime calculations as a function of the hypersperical basis size $K_{\max}$.}
\label{fig:hh-conv}
\end{figure}

The convergence of the three-body lifetime calculations is shown in Fig.\ \ref{fig:hh-conv}. We use fully dynamic three-body calculations up to $K_{\max}=22$, while for the larger $K_{\max}$ values, the basis size is reduced to $K=22$ using the adiabatic procedure (``Feschbach reduction'', see Ref.\ \cite{Grigorenko:2009}). It can be found that converged lifetime values are obtained for $E_r>2.2$ MeV at $K=60$. A value of $K \sim 100$ is required for convergence for $E_r>1.7$ MeV. For lower $E_r$ values calculations are not converged: the decay dynamics is changing to sequential decay and HH method is not suited for such situations. So, for this range of $E_r$ we therefore use extrapolations by IDDM curves.

\textit{Lifetime calculations.}
%
%
--- The calculations with potential P3 of this work provide considerably larger three-body decay widths that with P2, used in \cite{Grigorenko:2003b}, see Fig.\ \ref{fig:hh-conv} and Table \ref{tab:struct}. This can be explained by larger orbital size for this potential, see Fig.\ \ref{fig:res-wfs}, and consequently by larger two-body decay width, see Table \ref{tab:widths}. With potential P4 we obtain $2p$ width values which are about a factor of 2 larger than with P3. This difference is consistent with the simple estimate via squared ratio of two-body widths in the $^{65}$Se-$p$ channel $[\Gamma_r(\mbox{P4})/\Gamma_r(\mbox{P3})]^2=2.33$. We consider P4 set as too unrealistic because of too large TES, and everywhere rely on P3. We would like to point out that a factor 2 increase of the widths predicted for P3 is possible with tolerable modification of the single-particle WF geometry, and can be regarded as a measure of theoretical uncertainty of our calculations. Fortunately, this factor 2 is not large enough to modify major conclusions of this work. Potential set P5 contains interactions in $l=3$ and thus provide structure of $^{67}$Kr with strong $p/f$ configuration mixing which is roughly consistent with shell-model structure predictions from  \cite{Goigoux:2016}.

\begin{table}[b]
\caption{Structure of the three-body WF $\Psi^{(+)}_{E_T}$ in the internal region in terms of  for valence protons. Weights of the shell-model-like configurations $[l_{j}^2]_0$ are given in percent. Last row shows shell-model (SM) predictions for valence nucleon configurations from \cite{Goigoux:2016}.}
\begin{ruledtabular}
\begin{tabular}[c]{ccccc}
Pot.  & $[p_{1/2}^2]_0$ & $[p_{3/2}^2]_0$ & $[f_{5/2}^2]_0$ & $[f_{7/2}^2]_0$ \\
\hline
P3  & 12.1 & 87.2 & 0.1 & 0.07 \\
P5  & 1.3 & 18.3 & 64.1 &  16.2   \\
SM  & 13.8 & 17.6 & 67.2 & 2.4 \\
\end{tabular}
\end{ruledtabular}
\label{tab:struct}
\end{table}

The results of the three-body lifetime calculations as function of energy $E_r$ are shown in Fig.\ \ref{fig:lifetime-ot-er} (b) by dashed curves. Because the lifetime calculations with $E_r<1.7$ MeV are not converged, extrapolations to small $E_r$ values using IDDM calculations are shown (solid curves). Conclusion here is the same as for IDDM: For $\sim 100\%$ [$p^2$] structure of $^{67}$Kr (P3 potential) the agreement with experimental lifetime can be obtained for a broad range of $E_r$ values, while for realistic $^{67}$Kr structure (potential P5, $\sim 18\%$ $[p^2_{3/2}]$) the agreement is possible only in a narrow range of ``transition'' $E_r$ values.

\begin{figure}[t]
\includegraphics[width=0.49\textwidth]{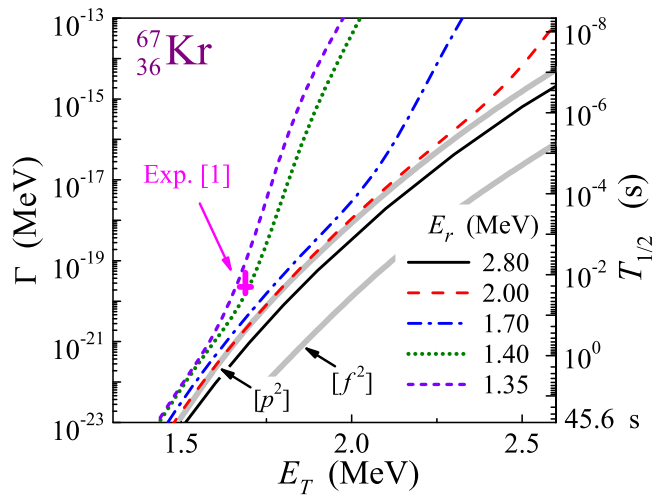}
\caption{(Color online) Lifetime of $^{67}$Kr as a function of $2p$ decay energy $E_T$ for several $E_r$ values. The three-body model with realistic structure (P5 potential) and IDDM extrapolation were used. The results of the three-body model calculations with pure $[p^2]$ and pure $[f^2]$ configurations from \cite{Grigorenko:2003b} are shown by thick gray curves.}
\label{fig:kr-lifetime}
\end{figure}

The lifetime predictions for $^{67}$Kr as a function of the total decay energy $E_T$ are given in Fig.\ \ref{fig:kr-lifetime} for different $E_r$ values. Variation of $E_r$ is obtained by varying the charged sphere radius $r_{\text{sph}}$ in the core-$p$ Coulomb potential. These are calculations with realistic $^{67}$Kr structure (P5 potential), performed at $E_T<0.7E_r$ (where they are reliably converged) and extrapolated to higher $E_T$ values using the IDDM curves. Transition from true $2p$ decay lifetime systematic to sequential $2p$  decay systematics can be seen for different $E_r$ values.

\begin{figure}[t]
\includegraphics[width=0.43\textwidth]{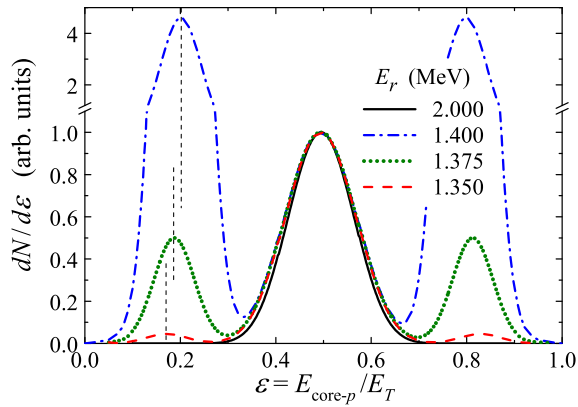}
\caption{(Color online) Energy correlations between core ($^{65}$Se) and one of the protons in $2p$ decay of $^{67}$Kr with $E_T=1.69$ MeV. Calculations by IDDM with different g.s.\ resonance energies $E_r$ in core-$p$ subsystem $^{66}$Br illustrate true $2p$ decay mechanism ($E_r=2$ MeV) and the transition decay dynamics region ($E_r=1375-1400$ keV). The curves are normalized to unity maximum value of the $\varepsilon \sim 0.5$ peak. The sequential decay peaks are convoluted with Gaussians with 150 keV FWHM. The ``low $\varepsilon$'' sequential decay peaks are indicated by vertical dashed lines.}
\label{fig:en-dis-y}
\end{figure}


\textit{Three-body correlations.}
%
%
--- As we have seen the situation with separation energies in $^{67}$Kr is very uncertain and there is a considerable chance that in this nuclide we face yet another example of transition dynamics (we can probably exclude possibility of pure sequential decay mechanism, as this immediately leads to very short lifetimes). The answer to the question about decay mechanism can be obtained by studies of energy correlations between core and one of the protons. Fig.\ \ref{fig:en-dis-y} shows these correlations in the case of true $2p$ decay and in the case of transition decay dynamics (transition ``true $2p$ decay'' $\rightarrow$ ``sequential $2p$ decay''). This transition is taking place in a very narrow interval of possible $^{66}$Br g.s.\ energies $E_r=1.35-1.42$ MeV. The estimated width of the $^{66}$Br g.s.\ is very small (e.g.\ $\Gamma_r= 1.17$ eV for $E_r=1.36$ MeV), therefore in Fig.\ \ref{fig:en-dis-y} we convolute sequential decay peaks with Gaussians of 150 keV FWHM for the sake of visual comparison. The probability of sequential decay changes from $\sim 5\%$ at $E_r=1.35$ MeV to $\sim 95\%$ at $E_r=1.40$ MeV together with more than order of the magnitude change in the lifetime (see Fig.\ \ref{fig:lifetime-ot-er}). This effect would evidently be observable for modern experiments with time projection chambers \cite{Miernik:2007b,Ascher:2011}.


\textit{Conclusions.}
%
%
--- The discrepancy between the lifetimes predicted for $^{67}$Kr in 2003  \cite{Grigorenko:2003b} and found in the recent measurements \cite{Goigoux:2016} inspired us to revisit the issue. We have reached the following conclusions:

\noindent (i) Various systematic studies favor for $^{67}$Kr either small positive or, which seems to be more probable, small negative proton separation energy $S_p$.

\noindent (ii) The experimentally observed lifetime of $^{67}$Kr can be explained by the true $2p$ decay mechanism, assuming dominance of $[p_{3/2}^2]_0$ configuration in the structure of $^{67}$Kr. This will work only if the $^{66}$Br g.s.\ is located close to or somewhat within the three-body decay ``energy window'', namely if $E_r=1.45-2.0$ MeV ($S_p$ from $-240$ to $310$ keV).

\noindent (iii) If we take into account the realistic structure of $^{67}$Kr, predicted in \cite{Goigoux:2016} with just $\sim 18\%$ of $[p^2_{3/2}]_0$ configuration, then the only possible way to explain the lifetime is to consider different decay mechanism. For $E_r=1.35-1.42$ MeV ($S_p$ from $-340$ to $-270$ keV) the decay of $^{67}$Kr corresponds to a ``transition dynamics'' on the borderline between true $2p$ and sequential $2p$ decay mechanisms. Further decrease of $E_r$ leads to pure sequential decay mechanism with rapid decrease of the lifetime beyond the experimentally acceptable value [see Fig.\ \ref{fig:lifetime-ot-er} (a)].

\noindent (iv) The question about the decay mechanism of $^{67}$Kr can be clarified by studies of $2p$ correlations. Predicted effects are strong enough to be observable by modern correlation experiment, even for very modest counting rates.

%
%
%


\bibliographystyle{apsrev}
\bibliography{d:/latex/all}


\end{document}